\documentclass[journal,article,submit,pdftex,moreauthors]{Definitions/mdpi} 
\let\linenumbers\relax
\usepackage{CJK}
\usepackage{amsmath}

\newcommand{\Msun}{M_\odot}

\newcommand{\Prad}{P_\mathrm{r}}
\newcommand{\Pgas}{P_\mathrm{g}}
\newcommand{\Pturb}{P_\mathrm{turb}}

\def\bfnabla{{\mbox{\boldmath $\nabla$}}}

\newcommand\bv{{\mbox{\boldmath $v$}}}

\newcommand\bn{{\mbox{\boldmath $n$}}}

\newcommand\bF{{\mbox{\boldmath $F$}}}

\newcommand{\hp}{H_{\mathrm{P}}}

\firstpage{1} 
\pubvolume{1}
\issuenum{1}
\articlenumber{0}
\pubyear{2022}
\copyrightyear{2022}
\datereceived{} 
\dateaccepted{} 
\datepublished{} 
\hreflink{https://doi.org/} 

\Title{Three Dimensional Natures of Massive Star Envelopes}

\TitleCitation{Three Dimsional Natures of Massive Stars Envelopes}

\Author{Yan-Fei Jiang \orcidA{}$^{\dagger}$}

\AuthorNames{Yan-Fei Jiang}

\AuthorCitation{Jiang, Yan-Fei}

\address{%
\quad $\dagger$Center for Computational Astrophysics, Flatiron Institute, 162 Fifth Avenue, New York, NY 10010, USA; yjiang@flatironinstitute.org}

\abstract{We review our current understanding on the outer envelope structures of massive stars based on three dimensional (3D) radiation hydrodynamic simulations. We briefly summarize the fundamental issues to construct hydrostatic one dimensional (1D) stellar evolution models when stellar luminosity approaches the Eddington value. Radiation hydrodynamic simulations in 3D covering the mass range from $13M_{\odot}$ to $80M_{\odot}$ always find a dynamic envelope structure with the time-averaged radial profiles matching 1D models with an adjusted mixing length parameter when convection is subsonic.  Supersonic turbulence and episodic mass loss are generally found in 3D models when stellar luminosity is super-Eddington locally due to the opacity peaks and convection is inefficient. Turbulent pressure plays an important role in supporting the outer envelope, which makes the photosphere more extended than predictions from 1D models. Massive star lightcurves are always found to vary with a characteristic timescale consistent with the thermal time scale at the location of the iron opacity peak. The amplitude of the variability as well as the power spectrum can explain the commonly observed stochastic low frequency variability of mass stars observed by TESS over a wide range of parameters in the HR diagram. The 3D simulations can also explain the ubiquitous macro-turbulence that is needed for spectroscopic fitting in massive stars. Implications of the 3D simulations for improving 1D stellar evolution models are also discussed.   }


\begin{document}




\section{Introduction}

Stellar evolution models in one dimension (1D) have been very successful to advance our understanding and make reliable predictions for properties of low mass stars \cite{Schalleretal1992,Paxton2011,Ekstrometal2012,Paxton2013,Paxton2015,Choietal2016,Paxton2018,Paxton2019,Jermynetal2023}, where radiation pressure is less important. However, uncertainties in the 1D models are significantly increased when massive stars (initial mass $\gtrapprox 8 M_{\odot}$) are considered. The remnant mass and properties of massive star envelopes, which directly determine their locations in the Hertzsprung–Russell (HR) diagram,  can vary significantly with  different assumptions adopted in 1D stellar evolution models \cite{Yusofetal2013,Kohleretal2015,Agrawaletal2022a,Agrawaletal2022}. 

One uncertainty is how to treat convection in massive star envelopes for 1D models. Convection can develop in massive star envelopes with enhanced opacity due to iron and helium \cite{Cantiello2009,Paxton2013}.  Convective energy transport in 1D is typically modeled based on the mixing length theory (MLT, \cite{Bohm-Vitense1958}), which assumes the stellar profile is in hydrostatic equilibrium and  calculates the energy flux entirely based on superadiabaticity using  local  fluid conditions (such as density, temperature) \cite[see the most recent review][]{JoyceTayar2023}. 
It gives fairly good results in the deep interiors of stars where convection is nearly adiabatic and very subsonic, although the results have some dependence on the assumed mixing length parameter. For convection in the envelope where the total optical depth to the photosphere (typically $\lesssim 10^4$) is not too large,  convective motion can be very inefficient to carry energy as radiative losses are not negligible. Modifications to the standard MLT in this regime are possible \cite{Henyey1965}, but it introduces an additional parameter that needs to be calibrated. Hydrostatic equilibrium is also commonly adopted in 1D models, which implies that the typical speed in convection should be much smaller than the sound speed. However, this is not always true for the convective speed returned by MLT formula in massive star envelopes \cite{Paxton2013}. This naturally raises a question of consistency in the treatment of convection. 

Another uncertainty is the treatment of radiation pressure, which can be the dominant pressure component in massive star envelopes. Convection near the photosphere is naturally very inefficient to carry energy as convective velocity is typically smaller 
than photon diffusion speed.   This implies that  stellar luminosity generated by the core  has to be transported by photons directly. This can cause the direct radiation acceleration to be larger than the gravitational acceleration in the regions where opacity is enhanced compared with the electron scattering value  in the envelope. This is a big issue for hydrostatic models as density would need to increase with radius to balance the excess radiation force. This will significantly change the predictions of stellar radii and photosphere temperatures in 1D models and cause severe numerical issues to evolve the star in 1D  \cite{Kohleretal2015,Yusofetal2013,Agrawaletal2022a,Agrawaletal2022}. For these radiation pressure-dominated flows, various radiation hydrodynamic instabilities distinct from convection can exist due to the density and temperature dependence of the opacity near composition-specific opacity peaks\cite{Kiriakidis1993,BlaesSocrates2003}. In many
cases, these cannot be captured by 1D stellar evolution models.  These
instabilities have been suggested as the origin of the large mass loss
rates from massive stars that exceed the value predicted by line-driven winds 
\cite{Glatzel1993}. Typical calculation of line-driven winds, which focuses on the optically thin region for the continuum radiation field, adopt a static and uniform envelope condition to begin with. Any clumping in the wind is generated by the instability of line-driven wind itself, which can already significantly affect the mass loss rate \cite{Owocki2015}. Supersonic convection in the envelope will very likely generate large amplitude density and velocity fluctuations of the photosphere, which can significantly affect the mass loss rate driven by the line force. It is unclear how good the Sobolev approximation \cite{Sobolev1960} is with this strongly turbulent photosphere. All of these point to another big uncertainty of massive stars, which is the mass loss rate that can significantly change the remnant mass \cite{Vink2022}.  Giant eruptions, or outbursts as in luminous blue variable stars may also play an important role in the evolution of massive stars. Their physical origins are likely coming from the interactions between the strong continuum radiation field and structures of these stars as reviewed by \cite{Davidson2020}.

Various multi-dimensional simulations have been performed in the recent years to study the outer envelope structures of massive stars, or the properties of line-driven winds in the optically thin part based on the CAK formula \cite{Freytag+2008,Chiavassa+2010,Jiang2015,Freytag+2017,Chiavassa+2018,Jiang2017,Jiang2018,Sundqvist+2018,Schultz2020,Moens+2022,Moens+2022b,Goldberg2022,Schultz+2023}. 
It is not trivial to map 3D stellar structures to 1D models, as turbulence in 3D can cause spatial and temporal correlations between different radiation-hydrodynamic quantities. Therefore, the commonly used quantities such as density, temperature, velocity, and radiation flux in 1D models cannot be simply mapped to volume-averaged 3D quantities. One example is the so-called "porosity" where density fluctuations typically anti-correlates with fluctuations of radiation flux, causing more photons to go through the low-density region \cite{Shaviv1998,Shaviv2000,Owocki+2004}, which effectively reduces the radiation force on the gas. However, these effects depend on the detailed properties of the turbulence and flow conditions (such as optically depth) \cite{Owocki+2018,Schultz2020}. Efforts to improve 1D models based on 3D simulations for massive star envelopes will also be discussed here.

The numerical complexity for ab initio simulations of massive star envelopes largely comes from the accurate treatment of radiation transport, which needs to account for both energy and momentum couplings between photons and gas, as well as a wide range of optical depth conditions. Diffusion approximation is widely adopted in 1D stellar evolution models and some 3D calculations \cite{Paxton2013,Moens+2022}, which is computationally efficient and a good approximation for the interior structures of stars.  However, the stellar envelope naturally includes the photosphere, including the transition region from optically thick to the optically thin part, where diffusion approximation is known to fail. Numerical algorithms for direct solver of the full radiation transport equation are available\cite{Jiang2012,Davis2012,Jiang2014,TsangMilosavljevic2015,Jiang2021,Jiang2022,Menon+2022}, which are computationally expensive. There is no direct comparison between different numerical schemes for problems specific to massive star envelopes yet. 
Opacity, which determines the interaction rates between photons and gas, is another big uncertainty for these calculations. Rosseland and Planck mean opacities are widely adopted in frequency-integrated radiation transport simulations, which are good approximations for very optically thick conditions. However, when optical depth drops, radiation force due to numerous lines becomes more and more important. Rosseland and Planck mean opacities are no longer able to correctly determine the momentum and energy couplings. For example, flux mean should replace Rosseland mean and they can be very different in the optically thin region\cite{Sander+2020,Vink2022}. However, flux mean opacity itself is generally not known as it depends on the solution to the transport equation. A review of current efforts for accurate radiation hydrodynamic simulations and prospects of future development will be provided.

\section{1D stellar evolution models}

We briefly summarize the main assumptions and equations solved for 1D stellar structures and highlight the issues related to massive star envelopes. The details can be found in textbooks \cite{Cox1968} and a recent review \cite{JoyceTayar2023}.  In this review, the envelope is defined as the region from the bottom of the convective zone due to iron opacity peak to the stellar surface. This typically covers the region with optical depth smaller than $10^4-10^5$ and temperature smaller than a few times $10^5$ K. Stellar winds are also included as part of the envelope, although we will not review the theory of mass loss for massive stars here as it has been done recently \cite{Davidson2020,Vink2022}. 

Structures of massive star envelopes in 1D are commonly constructed assuming hydro-static equilibrium and a constant luminosity, which can be expressed as
\begin{eqnarray}
	\frac{dP_g}{dr}+\frac{dP_r}{dr}&=&-\rho \frac{GM}{r^2},\nonumber\\
	L&=&4\pi r^2\left(F_r+F_c\right).
\end{eqnarray}
Here $P_g$ and $P_r$ are gas pressure and radiation pressure respectively, $r$ is the stellar radius, $G$ is the gravitational constant. The total mass $M$ inside $r$ is typically a constant for the envelope of main sequence stars, but not necessary for giants. The total luminosity $L$ is transported by the radiative flux $F_r$ and the convective flux $F_c$.  Gas and radiation are always assumed to be in thermal equilibrium so that gas temperature and radiation temperature are the same. Radiative flux is calculated based on gas temperature gradient via the diffusion approximation while the convective flux is determined based on MLT as described below. 

\subsection{Convective Flux based on Mixing Length Theory}
\label{sec:mlt}
The characteristic length scale under hydrostatic equilibrium is the pressure scale height $\hp$, defined as 
$\hp\equiv |P_t/\left(dP_t/dr\right)|=(P_g+P_r)r^2/(GM\rho)$, where total pressure is $P_t\equiv P_r+P_g$. For given radial profiles of total pressure and temperature, we can define an averaged temperature gradient with respect to the total pressure gradient as $\nabla\equiv d \ln T/d\ln P_t$. With convective eddies rising and falling in the star, the temperature change in the fluid elements with respect to the pressure change is defined as $\nabla^{\prime}$. In the deep interior of the star, it is commonly assumed that the fluid elements move adiabatically so that $\nabla^{\prime}=\left(dT/dP_t\right)_{\rm ad}$. The radiative flux is always calculated based on the diffusion equation as
\begin{eqnarray}
	F_r=-\frac{c}{\kappa_t\rho}\frac{dP_r}{dr}\approx \frac{ca_r}{3\kappa_t\rho}\frac{T^4}{\hp}=\frac{4ca_rT^4}{3\kappa_t\rho}\frac{\nabla}{\hp},
\end{eqnarray}
where $c$ is the speed of light, $a_r$ is the radiation constant while $\kappa_t$ is the total Rosseland mean opacity. In massive star envelopes when $P_r\gg P_g$, the pressure scale height $\hp$ is determined by the radiation pressure  and $\nabla=1/4$. We can also define a radiative gradient if we require that the total flux is only transported by radiation 
\begin{eqnarray}
	\frac{L}{4\pi r^2}\equiv \frac{4ca_rT^4}{3\kappa_t\rho}\frac{\nabla_r}{\hp}.
\end{eqnarray}
In the non-convective zone, $\nabla_r=\nabla$ and will be larger than the actual gradient when part of the luminosity is transported by convection. In general, we have the following relations in the envelopes $\nabla_r > \nabla > \nabla^{\prime} > \nabla_{\rm ad}$. Convective flux is determined by the typical length a convective eddy will travel $\lambda$, the convective velocity $\bar{v}$, the difference between the two gradients $\nabla, \nabla^{\prime}$ as \cite{JoyceTayar2023}
\begin{eqnarray}
	F_c=\frac{1}{2}\rho \bar{v} c_pT\frac{\lambda}{H_p}\left(\nabla-\nabla^{\prime}\right),
\end{eqnarray}
where $c_p$ is the heat capacity. Except for the convective velocity, which will be estimated below, all the other quantities in this equation are local conditions given by the background equilibrium profile. This is another implicit assumption of MLT, where the difference between $\nabla$ and $\nabla^{\prime}$ is believed to be small and each turbulent eddy typically travels a distance $\lambda$ that is smaller than the scale height $\hp$.

\subsection{Convective Velocities}
Convective velocity can be estimated based on the net acceleration for each eddy when it travels the distance $\lambda$ as \cite{JoyceTayar2023}
\begin{eqnarray}
	\frac{1}{2}\rho \bar{v}^2 =-\frac{1}{16} g\Delta\rho\lambda,
\end{eqnarray}
where $\Delta \rho$ is the average density contrast between each eddy and the background density profile. 
The density contrast can be related to the temperature difference via the thermal-dynamic properties of the gas and the turbulent velocity can be expressed as
\begin{eqnarray}
	\bar{v}  =  \frac{1}{2 \sqrt{2}} \lambda \, \bigg(\frac{g\, Q}{\hp}\bigg)^{1/2} \, (\nabla - \nabla^{\prime})^{1/2}
=  \frac{1}{2 \sqrt{2}} g \, \lambda \bigg(\frac{\rho\, Q }{P_t}\bigg)^{1/2}  \, (\nabla - \nabla^{\prime})^{1/2},
\label{eq:conV}
\end{eqnarray}
where $	Q = (4-3 \beta)/\beta - (\partial \ln \mu/\partial \ln T)_{P_t}$ and $\beta\equiv P_g/P_t$. Notice that the convective velocity is proportional to $Q^{1/2}$ and $Q$ will diverge when $\beta\rightarrow 0$ in the strongly radiation pressure-dominated regime. The ratio $\bar{v}/\sqrt{P_t/\rho}$ is also proportional to $\big[\lambda/\hp\big]Q^{1/2}(\nabla - \nabla^{\prime})^{1/2}$. This ratio may exceed unity in the radiation pressure-dominated regime, in particular if convection is inefficient and not able to bring $\nabla$ close to $\nabla^{\prime}$.

In the deep interior of stars, $\nabla^{\prime}$ is typically taken to be the adiabatic value. This is not necessarily the case in the envelope as photon diffusion time across $\lambda$ may be comparable or even smaller than the time each turbulent eddy takes to cross the mixing length $\lambda$. 
The efficiency of convective energy transport is determined by comparing $\bar{v}$ and photon diffusion speed $c/\tau_{\lambda}$, where $\tau_{\lambda}$ is the total optical depth across $\lambda$ in the convective zone. 
A general framework to include radiative losses of convective eddies in 1D models is described in \cite{Henyey1965}. Convective efficiency changes smoothly from opaque to transparent regimes. It has also been realized that turbulent pressure can become significant compared with the thermal pressure in this region and they should be included in the MLT formula \cite{Henyey1965}.

\subsection{Typical 1D models of Massive Star Envelopes}
\label{sec:1d_model}

\begin{figure}[htp]
	\centering
	\includegraphics[width=0.5\hsize]{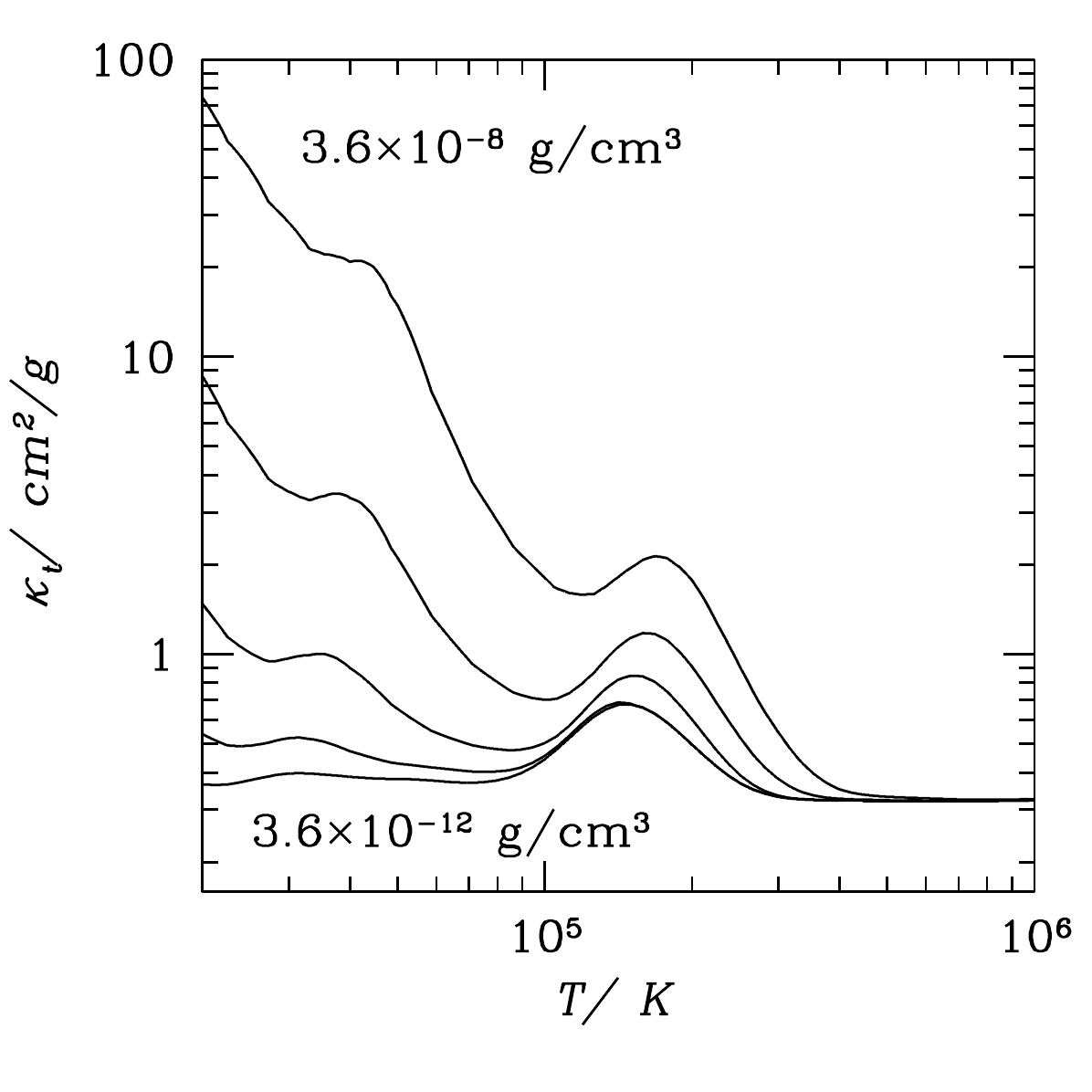}
	\caption{Rosseland mean opacity $\kappa_t=\kappa_s+\kappa_a$ at solar metallicity as a function of density and temperature, which is taken from Figure 2 of \cite{Jiang2015}. Each line shows the opacity variation with temperature for a fixed density value as indicated in the figure. Density decreases by a factor of 10 for each line from $3.6\times 10^{-8}$ g/cm$^3$ at the top to $3.6\times 10^{-12}$ g/cm$^3$ at the bottom.}
	\label{fig:opacity}
\end{figure}

 Massive star envelopes are radiative except the outer $2\%-4\%$ in radius, where isolated convective zones exist due to the enhanced iron opacity peak around temperature $2\times 10^5$ K and helium opacity around a few $10^4$ K, as shown in Figure 1 of  \cite{cantiello+2021} for the 1D envelope structures for a $20M_{\odot}$ star during the main sequence.  The typical Rosseland mean opacity as a function of density and temperature at solar metallicity is shown in Figure \ref{fig:opacity}. It can be shown \cite{Joss+1973,Paxton2013} that once the Eddington ratio, which is the ratio between local radiative acceleration and gravitational acceleration is larger than $\left(1-P_g/P_t\right)\left(\partial \ln P_r/\partial \ln P_t\right)_s$, convection will develop. Since luminosity and enclosed mass are constant in the envelope, the Eddington ratio, and therefore the convective zones, are tightly connected to the local opacity peaks. The convective zone caused by the iron opacity peak is typically located at the optical depth $\approx10^3- 10^4$ and will move deeper when the star evolves. The convective zone caused by the helium opacity peak is typically just below the photosphere and may not exist when the effective temperature is too high. Since hydrostatic equilibrium is always assumed in 1D models, it can be shown \cite{Paxton2013} that density will start to increase with radius when the Eddington ratio is larger than $1/\left[1+\left(\partial P_g/\partial P_r\right)_{\rho}\right]$, which is commonly referred as density inversion in the literature. These sub-surface convection zones are typically very inefficient, which means energy flux carried by convection is much smaller than the radiative flux as photon diffusion speed is larger than the typical convective speed as estimated based on equation \ref{eq:conV}.

Density inversion can cause big numerical trouble for stellar evolution as the time step often needs to be reduced significantly to get reasonable 1D solutions. Various numerical schemes have been used to pass this evolution phase, which results in very different envelope structures. In PARSEC (PAdova and TRieste Stellar Evolution code) models, the temperature gradient is limited to a maximum value set by hand so that density inversion never develops \cite{Alongi+1993,Chen+2015}. Similarly, the MLT++ formalism adopted in MESA \cite{Paxton2013} also reduces the temperature gradient as returned by MLT whenever the Eddington ratio exceeds a pre-defined threshold. Both methods effectively limit the radiative flux (therefore the radiation force) in the sub-surface convective zones, while the rest of the luminosity is transported via convection even though convection should be inefficient. This typically results in a more compact envelope. The Geneva models use a different approach \cite{Ekstrom2012}. They assume the mixing length $\lambda$ is proportional to the density scale height instead of the commonly used pressure scale height $\hp$ for stars more massive than $40M_{\odot}$. Since the density scale height is typically much larger than the pressure scale height in the radiation pressure-dominated envelopes, this can also enhance the convective flux and avoid the density inversion. Another completely different treatment is adopted by the BoOST (Bonn Optimized Stellar Tracks) models\cite{Szecsi+2022}. They do not include any enhancement of convective energy transport and therefore the envelope is inflated with a density inversion when the Eddington limit is reached 
\cite{Sanyal2015}. Further evolution of these models typically requires mass loss prescriptions to remove the outer layer of the star.  Envelope inflation is a natural solution to the hydrostatic equation near the Eddington limit \cite{Grafner+2012}. This is essentially because a very small density gradient is needed to maintain hydrostatic equilibrium when gravity is mostly balanced by radiation force. Because optical depth is proportional to $\rho\kappa$, the photosphere will not be reached until density drops to a small enough number at a large radius. 

These different 1D stellar evolution models can predict maximum stellar radii that differ by more than $1000R_{\odot}$ for stars between $40M_{\odot}$ and $100M_{\odot}$\cite{Agrawaletal2022a}. Even though stellar luminosity is similar among these models, the significant difference in stellar radii can cause the effective temperature to differ by a few thousand $K$\cite{Kohleretal2015}.  This can also cause the ionizing photons emitted by a given synthetic population to differ by $18\%$ \cite{Agrawaletal2022a} among the predictions based on different stellar evolution models.  Even though these models adopt very similar mass loss recipes, the actual mass loss rates are also very different as  photosphere conditions differ significantly. Therefore, remnant mass predicted by these models can differ by $\approx 10-20M_{\odot}$ for the same initial mass above $40M_{\odot}$.

\section{Structures of Massive Star Envelopes in 3D}
Convection is intrinsically a 3D phenomenon. In this section, we will first describe the general radiation hydrodynamic equations we solve in 3D and show how these equations are connected to the 1D prescriptions discussed in the literature. Then we will summarize the key results found by 3D simulations on properties of massive star envelopes and discuss possible ways to include these results in 1D stellar evolution models.

\subsection{Numerical Methods}
\label{sec:method}
Envelopes of massive stars extend from the photosphere to the deeper region with optical depth up to $\approx 10^3-10^4$, where the iron opacity peaks are located. We cannot assume an adiabatic equation of state for the gas and radiation transport is needed to determine the thermal properties of gas self-consistently. Moreover, for a typical density range $\approx 10^{-10}-10^{-8}$ g/cm$^3$ and temperature around $10^5$ K in the envelopes, radiation pressure can typically dominate over gas pressure. Therefore, both energy and momentum exchanges between photons and gas need to be included  to model the dynamics of the envelopes correctly. Near the photosphere, traditional diffusion approximation does not apply and the full transport equation needs to be solved. Given all these considerations, the following set of radiation hydrodynamic equations are typically solved \cite{Jiang2021}

\begin{equation} \label{eq:RHD}
\begin{split}
\frac{\partial\rho}{\partial t}+\bfnabla\cdot(\rho\bv)&=0,\\
\frac{\partial(\rho\bv)}{\partial t}+\bfnabla\cdot\left(\rho\bv\bv+P_g\mathbb{I}\right) &=-\mathbf{G}_r-\rho\bfnabla\Phi, \\
\frac{\partial{E}}{\partial t}+\bfnabla\cdot\left[(E+P_g)\bv\right]&=-c G^0_r -\rho\bv\cdot\bfnabla\Phi,\\
\frac{\partial I}{\partial t}+c\bn\cdot\bfnabla I&=S(I,\bn),
\end{split}
\end{equation}
 where $\rho,\bv,P_g,\Phi$ are gas density, velocity, pressure and gravitational potential respectively while $\mathbb{I}$ is the identity tensor. The total energy $E$ includes gas internal and kinetic energy as $E\equiv \rho v^2/2+P_g/(\gamma-1)$, where $\gamma$ is the adiabatic index of the gas. The radiation field is described by the frequency integrated lab frame specific intensity $I$, which is a function of time $t$, spatial location, and photon propagation direction with unit vector $\bn$. The commonly used radiation energy density $E_r$, radiation flux $\bF_r$ and radiation pressure ${\sf\Prad}$ are just angular quadrature of specific intensities as
\begin{equation}
E_r=\int I d\Omega, \bF_r=c\int I\bn d\Omega, {\sf \Prad}=\int \bn\bn I d\Omega,     
\end{equation}
where $\Omega$ is the weight of angular quadrature for each angle $\bn$ and $c$ is the speed of light. The radiation energy $(G^0_r)$ and momentum ($\mathbf{G}_r$) source terms also correspond to the direct quadrature of the source term for specific intensities $S(I,\bn)$ so that total energy ($E+E_r$) and momentum ($\rho\bv+\bF_r/c^2$) are conserved in the lab frame during photon-gas interactions. The full expression for the source term $S(I,\bn)$ can be found in \cite{Jiang2021}. In the co-moving frame of the fluid, it includes isotropic scattering, emission and absorption as
\begin{eqnarray}
 S_0(I_0) = c\left[\rho(\kappa_a+\kappa_s)(J_0-I_0)+\rho\kappa_p\left(\frac{a_rT^4}{4\pi} - J_0\right)\right],
\end{eqnarray}
where $I_0$ is the co-moving frame specific intensity and it is related to lab frame value $I$ via the Lorentz transformation. We have also defined the mean intensity in the co-moving frame as $J_0$ and $\kappa_s$ is the scattering opacity, $\kappa_a$ is the Rosseland mean absorption opacity, $\kappa_p$ is the Planck mean opacity. These mean opacities can be calculated based on the monochromatic opacities as described below. The lab frame source term $S(I,\bn)$ is related to $S_0(I_0)$ with proper transformation as described in \cite{Jiang2021}. 

To simplify things further, moments of specific intensity are typically evolved and the governing equation for radiation energy density $E_r$ and radiation flux $\bF_r$  can be derived by integrating the specific intensities $I$ over angles $\bn$ and keeping only terms to $\mathcal{O}(v/c)$ and some $\mathcal{O}(v/c)^2$ terms as \citep{MihalasKlein1982,Lowrie1999,Jiang2012}
\begin{equation}
    \begin{split}\label{eq:rad_mom}
        \frac{\partial E_r}{\partial t}+\bfnabla\cdot\bF_r=cG_r^0=c\rho\kappa_p\left(a_rT^4-E_r\right)
        +\frac{\rho(\kappa_a-\kappa_s)}{c}\bv\cdot\bF_{r,0}, \\
        \frac{1}{c^2}\frac{\partial \bF_r}{\partial t}+\bfnabla\cdot {\sf \Prad}=\mathbf{G}_r=-\frac{\rho\left(\kappa_s+\kappa_a\right)}{c}\bF_{r,0}+\frac{\rho\kappa_p}{c} \left(a_rT^4-E_r\right)\bv,
    \end{split}
\end{equation}
where the co-moving frame radiation flux $\bF_{r,0}$ is related to the lab-frame flux $\bF_r$ to the first order $v/c$ as $\bF_{r,0}=\bF_r-\bv\cdot\left(E_r\mathbb{I}+{\sf \Prad}\right)$.
It is now possible to solve the full time-dependent evolution of specific intensities $I$ directly (equation \ref{eq:RHD}) using the algorithm developed by \cite{Jiang2012}. The radiation moment equations \ref{eq:rad_mom} do not need to be solved directly with an appropriate closure scheme as done by \cite{Jiang2012,Davis2012}. But they are still useful for analysis of simulation results. Co-moving frame formula is also widely used for radiation transport in stellar envelopes, where the co-moving frame radiation energy density $E_{r,0}$ and flux $\bF_{r,0}$ are evolved as the fundamental quantities\cite{HillierMiller1998,Puls+2020}. This is a natural way to describe the interaction terms between radiation and gas as the opacity is defined in the gas rest frame, although this will complicate the advection term and the equation cannot be solved in a totally conservative format.

Notice that the radiation hydro equations given by equation \ref{eq:RHD} are completely general and the frequency dependence can also be included \cite{Jiang2022}. In principle, this can be used to model the radiation force due to lines in multi-dimensional simulations. 
However, it is still too expensive to resolve each line, or even groups of lines in the frequency space for full radiation hydrodynamic simulations.  In practice, frequency-integrated specific intensity $I$ is typically evolved, which requires properly weighted opacity. If we assume absorption and scattering processes are isotropic in the \emph{co-moving} frame, we can use the opacity weighted by radiation energy density and flux to approximate the intensity-weighted opacity. In principle, based on the radiation moment equation \ref{eq:rad_mom}, the opacities we need are
\begin{eqnarray}
	\kappa_F &\equiv& \frac{\int_{\nu} \left[\kappa_a(\nu)+\kappa_s(\nu)\right] F_{r,0}(\nu)d\nu}{\int_{\nu} F_{r,0}(\nu)d\nu}, \nonumber \\
	\kappa_p&\equiv &\frac{\int_{\nu} \kappa_a(\nu) B_r(\nu,T)d\nu}{\int \nu B_r(\nu,T)d\nu}, \nonumber\\
	\kappa_{E_r}&\equiv& \frac{\int_{\nu} \kappa_a(\nu)E_{r,0}(\nu) d\nu}{\int_{\nu} E_{r,0}(\nu)d\nu},
\end{eqnarray}
where $B_r(\nu,T)$ is the blackbody spectrum at temperature $T$ while $\kappa_a(\nu)$ and $\kappa_s(\nu)$ are the monochromatic absorption and scattering opacities, which are also a function of density and temperature in general. 
The averaged opacities $\kappa_F,\kappa_p,\kappa_{E_{r}}$ are typically referred to as flux mean, planck mean and energy mean. However, $\kappa_F$ and $\kappa_{E_r}$ depend on solutions to the transport equation themselves and are typically unknown in advance. In the optically thick regime, if we make the assumptions that the co-moving frame radiation energy density follows the blackbody spectrum and 
the diffusion equation holds for the co-moving frame radiation flux in the whole spectrum, we can use $\kappa_p$ to replace $\kappa_{E_r}$ and $\kappa_F$ can be replaced with the Rosseland mean opacity defined as
\begin{eqnarray}
	\kappa_R\equiv \left[\frac{\int_{\nu}\left(\partial B_r(\nu,T)/\partial T\right) \left[1/\left(\kappa_a(\nu)+\kappa_s(\nu)\right)\right]d\nu}{\int_{\nu}\partial B_r(\nu,T)/\partial T d\nu}\right]^{-1}.
	\label{eq:rossmean}
\end{eqnarray}
The opacities $\kappa_a$ and $\kappa_s$ used in equations \ref{eq:RHD} and \ref{eq:rad_mom} are the Rosseland mean absorption and scattering opacities, which need to be calculated together according to equation \ref{eq:rossmean}. Clearly, Rosseland mean opacities will only fail when either one of the assumptions is broken, which also means that when better opacity is considered in the transport  equation, diffusion approximation also needs to be improved to get a more accurate solution to the transport equation. 
 
Gravitational potential used in equation \ref{eq:RHD} is taken to be $-GM_c/r$, where $G$ is the gravitational constant and $M_c$ is the total mass inside radius $r$. We typically take $M_c$ to be a constant value when the envelope mass is much smaller than the core mass, which may not be the case for evolved red supergiant stars \citep{Goldberg2022}.

It is useful to average this set of equations over all $\theta$ and $\phi$ directions in the spherical polar coordinate and compare it with the commonly used 1D formula for stellar structures. For any quantity $a$, we define the shell averaged value as
\begin{equation}
\langle a\rangle\equiv \frac{\int_{\theta}\int_{\phi} a\sin\theta d\theta d\phi}{\int_{\theta}\int_{\phi} \sin\theta d\theta d\phi}.
\end{equation}
Then the shell-averaged continuity, momentum and energy equations are
\begin{equation} \label{eq:1Daverage}
\begin{split}
\frac{\partial \langle\rho\rangle}{\partial t}+\frac{1}{r^2}\frac{\partial }{\partial r}\left(r^2\langle\rho v_r\rangle\right)=0, \\
\frac{\partial \langle\rho v_r\rangle }{\partial t}+\frac{\partial \langle \rho v_r^2 \rangle}{\partial r}+\frac{\partial \langle \Pgas\rangle}{\partial r}+\frac{2\langle \rho v_r^2\rangle-\langle\rho v_{\theta}^2\rangle-\langle \rho v_{\phi}^2\rangle }{r}=-\langle G_{r,r}\rangle - \frac{G M_c\langle\rho\rangle }{r^2},\\
\frac{\partial \langle E \rangle}{\partial t}+\frac{1}{r^2}\frac{\partial }{\partial r}\left(r^2\langle \frac{1}{2}\rho(v_r^2+v_{\theta}^2+v_{\phi}^2)v_r+\frac{\gamma}{\gamma-1}\Pgas v_r\rangle\right) =-c\langle G_r^0\rangle-\frac{GM_c\langle \rho v_r\rangle }{r^2}. 
\end{split}
\end{equation}
Equations for shell-averaged radiation energy density and flux are
\begin{equation}
\begin{split}
\frac{\partial \langle E_r\rangle}{\partial t}+\frac{1}{r^2}\frac{\partial r^2\langle F_{r,r}\rangle }{\partial r}=c\langle G_r^0\rangle, \\
\frac{1}{c^2}\frac{\partial \langle F_{r,r}\rangle}{\partial t}+\frac{\partial \langle P_{r,r}\rangle }{\partial r}=\langle G_{r,r}\rangle.
\end{split}
\end{equation}
Here $v_r,v_{\theta},v_{\phi}$ are flow velocity along $r,\theta,\phi$ direction respectively while $F_{r,r}$ and $P_{r,r}$ are radial component of radiation flux and $r-r$ component of the radiation pressure tensor. To the first order of $v/c$, the radial component of radiation momentum source term 
is $\langle G_{r,r} \rangle\approx -\rho (\kappa_s+\kappa_a)F_{r,0,r}/c$, where $F_{r,0,r}$ is the radial 
component of the co-moving frame radiation flux. 

If the flow is static, these equations are reduced to the standard 1D equation for stellar structures, where total luminosity is $4\pi r^2F_{r,r}$ and gravitational force is balanced by $\partial \Pgas/\partial r+\partial P_{r,r}/\partial r$. Once turbulence develops, additional terms related to the flow velocity show up in the momentum and energy equations. If the turbulence is isotropic  so that 2$\langle \rho v_r^2\rangle=\langle\rho v_{\theta}^2+\rho v_{\phi}^2\rangle$, or the scale height is much smaller than radius so that 
$\partial\langle \rho v_r^2\rangle/\partial r \gg \langle2\rho v_r^2-\rho v_{\theta}^2-\rho v_{\phi}^2 \rangle/r$, then all the additional terms in the momentum equation can be treated as an effective pressure $\Pturb\equiv \rho v_r^2$. This is usually called turbulent pressure in the literature. If the turbulence is very subsonic, these terms can be safely neglected. However, for massive star envelopes,  supersonic turbulence can develop and then the velocity-dependent terms are important to support gravity as we will show in the following section. 

Energy flux is the superposition of diffusive flux $F_{r,0}$ and advective flux $\left\langle \left(E_r+P_{r,r}\right) v_r\right\rangle +\langle \rho\left(v_r^2+v_{\theta}^2+v_{\phi}^2\right)v_r/2+\gamma\Pgas v_r/(\gamma-1)\rangle$. For non-zero mass flux (such as wind from the envelope), the energy flux will also include the gravitational term $-\langle\rho v_r\rangle GM_c/r$. The advective term is essentially what the mixing length theory tries to calculate, which is typically called convective flux in that framework. Because of the turbulent nature of convection, it cannot be simply calculated by $\langle v_r\rangle$ multiplied by the shell averaged energy density. The relative importance of energy transport by radiative diffusion $\bF_{r,0}$ and convection then depends on the optical depth per pressure scale height in the convective zone. We will quantify this in the following section. 

\subsection{Convection in 3D}
Massive star envelopes can become convectively unstable because of the iron opacity peak as discussed in section \ref{sec:1d_model}, which is typically located around the temperature $\approx 1.8\times 10^5$ K with a relatively weak dependence on density as shown in Figure \ref{fig:opacity}. The Rosseland mean opacity can be a factor of $2-5$ times the electron scattering value, which can easily cause the radiation force to exceed the gravitational force if all the luminosity generated from the core is transported by diffusive radiation flux in the envelope. The envelope structure is entirely dependent on the outcome of convection. In particular, 3D simulations will check whether  the majority of luminosity will be carried by the convective term or not without any ad hoc assumptions.  The relative importance of convective and radiative fluxes will depend on the optical depth in the convective zone, which itself will vary as the star evolves in the HR diagram.

Since the iron opacity peak is located around the temperature $T_{\rm Fe}=1.8\times 10^5$ K, the optical depth $\tau_{\rm Fe}$ between its location and the photosphere can be roughly estimated as 
\begin{equation}\label{eq:tau_fe}
\tau_{\rm Fe}\approx \frac{2}{3}\left(\frac{T_{\rm Fe}^4}{T_{\rm eff}^4}-1\right),   
\end{equation}
where $T_{\rm eff}$ is the photosphere temperature. This estimate is based on diffusion approximation and assumes that diffusive radiation flux is a constant value $ca_rT_{\rm eff}^4/2$. For example, $\tau_{\rm Fe}\approx 7\times 10^4$ for $T_{\rm eff}=10^4$ K and it drops to $111$ when $T_{\rm eff}$ is increased to $5\times 10^4$. This suggests that when massive stars evolve away from the main sequence with a decreasing effective temperature, the iron opacity peak will move deeper to the envelope with an increasing $\tau_{\rm Fe}$. This simple estimate is consistent with 1D stellar evolution models \cite{Cantiello2009,Cantiello2021}.

Properties of convection in these radiation pressure-dominated massive star envelopes were first studied self-consistently by \cite{Jiang2015}, which solves the 3D radiation hydrodynamic equations in Cartesian geometry under plane parallel approximation using a closure scheme based on variable Eddington tensor (VET) \cite{Jiang2012,Davis2012}. Three models are chosen to cover the effective temperature range from $7.13\times 10^3$ K to $4.75\times 10^4$ K with luminosity varying from $4\times 10^5L_{\odot}$ to $1.3\times 10^6L_{\odot}$ based on MESA stellar evolution models of $80M_{\odot}$ and $40M_{\odot}$ stars. All these calculations start from initial profiles that are in hydrostatic equilibrium with a constant radiation flux.  The bottom boundary conditions and core mass are set to values given by MESA models. These profiles are intrinsically unstable to convection in 3D, which results in turbulence in the envelope. After a few thermal timescales of the envelope, it will settle down to a structure with self-consistent velocity fields. One important conclusion from comparing the three models is that envelope structures 
strongly depend on the optical depth per pressure scale height $\tau_0$ at the radius where the iron opacity peak is located as compared to a critical value $\tau_c$ defined as
\begin{equation}
    \tau_c=\left(\frac{P_r}{P_r+P_g}\right)c/\sqrt{k_BT_{\rm Fe}/(\mu m_p)}.
\end{equation}
For $T_{\rm Fe}=1.8\times 10^5$ K and mean molecular weight $\mu=0.6$, the critical optical depth is $\tau_c=6000$ and $\tau_0$ can be reasonably estimated by $\tau_{\rm Fe}$ (Table 1 of \cite{Jiang2015}). Since the diffusive flux can be estimated as $c \langle E_r \rangle /\tau_0$ while the convective flux is $\approx \langle 4v_rE_r/3 + 5P_gv_r/2 \rangle$ with the characteristic velocity being the gas sound speed, the comparison between $\tau_c$ and $\tau_0$ determines the relative importance of diffusive flux and convective flux for energy transport in the radiation pressure dominated envelope.

When $\tau_0\gg \tau_c$, convective flux is larger than diffusive flux, which typically happens for post main sequence stars. Since radiation force is only determined by the product of opacity and local diffusive flux, the Eddington ratio in the turbulent envelope is much smaller than the value estimated based on the total luminosity. 
Convective turbulence in the envelope is typically sub-sonic and density fluctuation is smaller than $10\% - 20\%$ 
of the mean values. This is also the regime where the mixing length type formula for convection could apply with properly adjusted parameters. 
By comparing the time and horizontally averaged convective flux from the simulation and MLT formula, Figure 6 of \cite{Jiang2015} shows that  MLT works reasonably well to describe the convective flux with $\alpha=0.55$ for the envelope of a post main sequence $40M_{\odot}$ star.

\begin{figure}[htp]
	\centering
	\includegraphics[width=0.8\hsize]{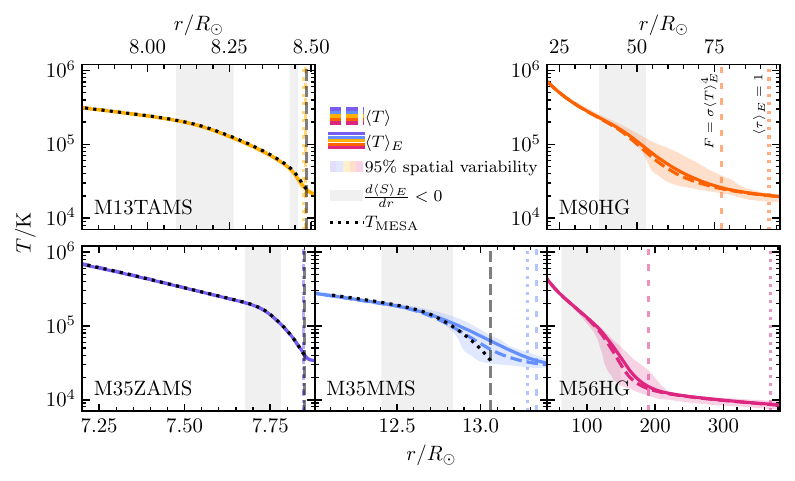}
	\caption{Comparison of radial profiles of properly averaged gas temperature in massive star envelopes from 3D radiation hydrodynamic simulations and 1D MESA profiles, adopted from Figure 3 of \cite{Schultz+2023}. The MESA profiles always stop at $\tau=2/3$. No MESA models are created to match M80HG and M56HG. The label for each run indicates the core mass and evolution stage as defined in \cite{Schultz+2023}. 
		The shaded color regions show the $95\%$ spatial variability due to convection in a single temporal snapshot.
		Vertical colored lines show the radii where the photosphere is in 3D based on two different definitions. 
		The grey shaded regions denote where the average entropy gradient is negative, indicating a convective zone.
		Temperature profiles from MESA models of similar stars to M13TAMS, M35ZAMS, and M35MMS are plotted in black dotted lines with their photospheric radii shown by the vertical grey dashed lines.}
	\label{fig:compare3D_1D}
\end{figure}

In the regime that $\tau_0 \ll \tau_c$, radiative diffusion dominates over convection for energy transport, which is typically the case for main sequence stars. There is no mysterious mechanism that can transport the energy without causing strong radiation force. In other words, convection is indeed inefficient as predicted by MLT for this regime. However, the envelope does not maintain a hydrostatic structure with a fixed density inversion. Instead, the envelope is very dynamic.  The initial density inversion in the hydrostatic profile is unstable to convection and some gas is accelerated to supersonic speed (with respect to gas sound speed). Shocks are formed in the envelope and large density fluctuations are produced. The iron opacity zone is initially pushed out by the strong radiation force. But all the gas falls back after photons escape and density inversion starts to develop again during this process. This causes the whole envelope to oscillate with the dynamical time scale at the convective zone. The time-averaged envelope structure still shows a modest density inversion (Figure 19 of \cite{Jiang2015}) for the zero-age main sequence model, which means the density fluctuation is not enough to reduce the Eddington ratio to be smaller than 1 with the porosity effect (see section \ref{sec:porosity}).

Most of the results are confirmed with improved simulations\cite{Jiang2017,Jiang2018,Schultz2020,Schultz+2023}, where the spherical polar coordinate is used so that geometric dilution and the gravitational potential can be properly captured, particularly for evolved stars where the pressure scale height can be a significant fraction of the stellar radius. As shown in Figure \ref{fig:compare3D_1D}, for 13$M_{\odot}$ main sequence star, or zero-age main sequence $35M_{\odot}$ star, radial profiles of shell averaged temperature profiles from 3D simulations agree with 1D MESA profiles very well. Convection is very subsonic and inefficient. No density inversion is formed as the Eddington ratio is smaller than one for the two cases.
For the middle age main sequence $35M_{\odot}$ and evolved $56M_{\odot}$ and $80M_{\odot}$ models, 3D simulations typically find the envelope is much more extended with the turbulent velocity exceeding the sound speed and convection is also inefficient. The radiative region below the convection zone still agrees with the 1D solutions very well but properties of the convective zone cannot be matched to the MLT formula described in section \ref{sec:1d_model}, as equation \ref{eq:conV} would predict a convective velocity much smaller than what is found by the simulations.  For stars with effective temperature below $10^4$ K, \cite{Jiang2018} shows that the extended envelope in 3D can cause significant opacity enhancement due to helium recombination for high density clumps formed in convection. This can cause significant episodic mass loss from the stellar surface with the time-averaged mass accretion rate reaching $\approx 10^{-7}-10^{-5} M_{\odot}\ {\rm yr}^{-1}$. The mass loss rate produced by this mechanism gets smaller when the effective temperature is higher as more helium is ionized. It is important to notice that this is different from the classical line-driven wind as the launching region is still very optically thick and it is entirely produced by the continuum radiation field with a local Eddington ratio significantly larger than unity. The mass loss is also not steady as the high density clumps are produced by the chaotic turbulent flow. However, it will be very interesting to include line force in these calculations to see how the wind properties will be modified. It is also interesting to notice that this significantly enhanced helium opacity is typically not found in 1D models. As we will explain in section \ref{sec:turbv}, it is the turbulent pressure that makes the envelope much more extended and significantly increases the density around the helium opacity zone, which cannot be captured by existing hydrostatic 1D models. 

\begin{figure}
	\centering
	\includegraphics[width=0.48\textwidth]{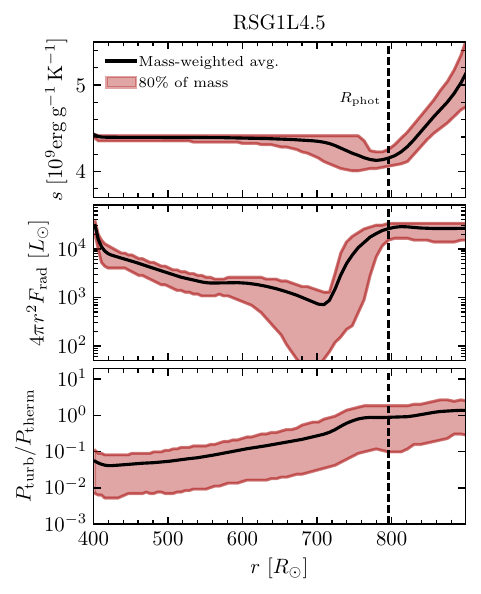}
	\includegraphics[width=0.48\textwidth]{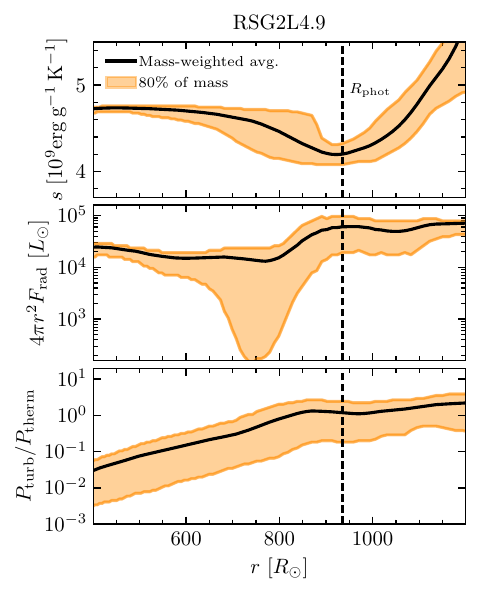}
	\caption{ \label{fig:rsg_profile} Radial profiles of volume averaged specific entropy (top panels), radiative luminosity (middle panels) and ratio of turbulent pressure to thermal pressure (bottom panels) from two 3D simulations of red supergiant stars taken from Figure 11 of \cite{Goldberg2022}. The left column is for RSG1L4.5 model with luminosity  $\log\left(L/L_{\odot}\right)=4.5$ at day 4707 while the right column is for RSG2L4.9 model with luminosity $\log\left(L/L_{\odot}\right)=4.9$ at day 4927. The vertical dashed lines indicate the location of photosphere.}
\end{figure}

For efficient convection in the envelope of red supergiant stars, 3D simulations find that the envelope structure 
can be very well described by the MLT formula. Figure \ref{fig:rsg_profile} shows radial profiles of various quantities from two red supergiant stars described by \cite{Goldberg2022}. When both radiation and gas entropy are included, specific entropy is a constant as a function of radius in the deep envelope as it should be for efficient convection. This is not the case for some earlier 3D simulations of red supergiant stars (for example, Figure 3 of \cite{Chiavassa+2011}), where radiation pressure is typically neglected. The radiative luminosity is only a few percent of the total luminosity in the deep convective zone but it always becomes the dominant energy transport mechanism near the photosphere. Turbulent pressure is negligible compared with thermal pressure in the deep envelope but they become comparable near the photosphere. Radial profiles of turbulent velocity as well as the superadiabaticity $\nabla-\nabla_{\rm ad}$ in the convective zone can be roughly matched to the a 1D solution based on MLT if  the dimensionless mixing length parameter $\alpha$ is chosen to be $3-4$\cite{Goldberg2022}. It is a little surprising that MLT can work in this case as MLT formula is entirely based on local conditions of the flow while convective eddies can have coherent structures across the whole star. They always show a topology of large area upwellings surrounded by narrow lanes of downward flows, which is also observed in 3D simulations of convection for solar type stars \cite{SteinNordlund1998}. Perhaps that is why the best calibrated mixing length parameter is much larger than $1$ in this case and it is smaller than $1$ when the size of the convective eddies is much smaller than the radius as shown in \cite{Jiang2015}.

\subsection{Turbulent Pressure Support}
\label{sec:turbv}
Turbulent velocity in the deep stellar interior (when $\tau > \tau_c$) is typically much smaller than the sound speed so that all the terms related to turbulent pressure $\langle\rho v_r^2\rangle,\langle \rho v_{\theta}^2\rangle,\langle \rho v_{\phi}^2\rangle$ in the momentum equation (second line of equation \ref{eq:1Daverage}) can be safely neglected compared with the gas and radiation pressure terms. However, near the photosphere, turbulent velocity can become comparable to, or even larger than the sound speed as found by many numerical simulations of convection \cite{Freytagetal1996,SteffenFreytag2005,trampedachetal2014}. This is particularly true for massive stars with significant radiation pressure support near the photosphere as discussed in previous sections.  The potential importance of turbulent pressure on the structure of massive stars has been realized for a long time \cite{deJager1984}. However, detailed studies of turbulent pressure and its consequence are only possible with 3D radiation hydrodynamic simulations.

\begin{figure}[htp]
\centering
\includegraphics[width=1.05\hsize]{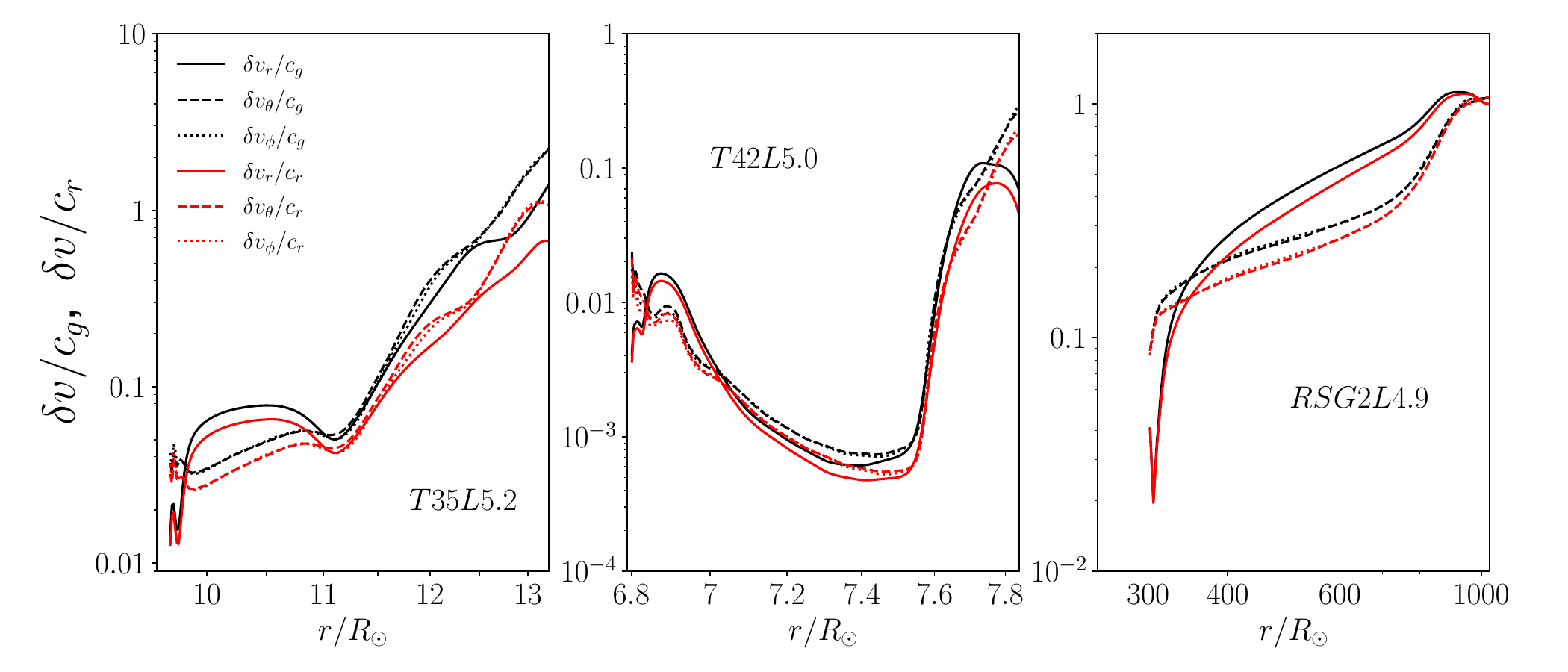}
\caption{Time averaged profiles of three components of turbulent velocities scaled with the local isothermal gas sound speed $c_g$ and radiation sound speed $c_r$ from two simulations $T35L5.2$ and $T42L5.0$ shown in \cite{Schultzetal2022} and $RSG2L4.9$ from \cite{Goldberg2022}. For each snapshot, we only show the region inside the photosphere.}
	\label{fig:turbulentpressure}
\end{figure}

Figure \ref{fig:turbulentpressure} shows radial profiles of turbulent velocities scaled with isothermal gas sound speed $c_g$ and radiation sound speed $c_r\equiv \sqrt{E_r/(3\rho)}$ for snapshots from three different simulations. The run $T35L5.2$ in the left panel is for a $35\Msun$ star halfway through the main sequence while the run $T42L5.0$ in the middle panel is for a zero-age main sequence (ZAMS) $35\Msun$ star \citep{Schultzetal2022}. These are the same simulations as labeled M35MMS and M35ZAMS by \cite{Schultz+2023} as well as in Figure \ref{fig:compare3D_1D}. The simulation $RSG2L4.9$ shown in the right panel is a $13\Msun$ mass red supergiant star described in \cite{Goldberg2022}. In each model, we only show the region inside the photosphere. Each component of the turbulent velocity $\delta v$ is calculated as 
$\delta v\equiv \left(\langle\rho (v - \langle v\rangle_{\rho})^2\rangle/\langle \rho \rangle\right)^{1/2}$, where $\langle v\rangle_{\rho}$ is the density weighted, shell averaged velocity. The turbulent velocities are always smaller than $10\%$ of the sound speed for the ZAMS model, where pressure scale height is only $1\%$ of the stellar radius and $\tau_{\rm Fe}\ll \tau_c$. For the model $T35L5.2$ where $\tau_{\rm Fe}$ becomes comparable to $\tau_c$ and pressure scale height becomes $2\%$ of radius, the turbulent velocities increase from a few percent of the sound speeds in the deep envelope where $\tau\approx 10^5$ to values that are larger than the sound speed at the photosphere. This is also true for the red supergiant model $RSG2L4.9$ where pressure scale height is comparable to the radius and $\tau_{\rm Fe}\gg \tau_c$. The whole envelope is convective for this model and the ratio between turbulent velocity and sound speed is typically larger than the ratio in the main sequence models. The three turbulent velocity components $\delta v_r,\delta v_{\theta}$ and $\delta v_{\phi}$ are comparable for $T35L5.2$ and $T42L5.0$ while the radial component is slightly larger than the horizontal components for $RSG2L4.9$ except near the photosphere. The strong turbulent pressure support makes the density scale height in 3D simulations much larger than the values found in 1D models. For lower mass stars or stars at the zero-age main sequence, turbulent pressure is typically less important near the photosphere.

When turbulent velocity exceeds the sound speed, it not only modifies the momentum equation and thus the hydrostatic equilibrium assumption but also impacts the energy equation. In particular, the temperature gradient $\nabla$ does not approach the adiabatic value $\nabla_{\rm ad}$ as commonly assumed in MLT. Instead, it should be the turbulent pressure modified value $\nabla^{\prime}_{\rm ad}$ as defined as \cite{Henyey1965}
\begin{eqnarray}
	\nabla^{\prime}_{\rm ad}\equiv \nabla_{\rm ad}\times \frac{d\ln P_t}{d\ln P_{\rm sum}},
\end{eqnarray}
where $P_t$ is the thermal pressure (gas plus radiation) while $P_{\rm sum}$ is the sum of $P_t$ and turbulent pressure $P_{\rm turb}$. The superadiabaticity should also be calculated as $\nabla-\nabla^{\prime}_{\rm ad}$, which can be used in the MLT formula. Figure 4 of \cite{Schultz+2023} shows that in the convective zones of 3D simulations, $\nabla$ is much closer to $\nabla^{\prime}_{\rm ad}$ instead of $\nabla_{\rm ad}$. Furthermore, the definition of convection zone, where specific entropy decreases as radius increases, agrees identically with the regions where $\nabla > \nabla^{\prime}_{\rm ad}$ instead of the traditional criterion $\nabla > \nabla_{\rm ad}$. 
Since turbulent pressure found by the simulations can be quite different from the values predicted by MLT, a self-consistent modification of MLT to account for the turbulent pressure will need a recipe to relate the thermal pressure and turbulent pressure based on the 3D simulations. 

 In 1D stellar evolution models, density scale height between the convective zone due to the iron opacity peak and the photosphere is typically very small, and density drops quickly with radius. For stars with effective temperature $\lesssim 2\times 10^4$ K, even though a narrow convective zone due to the helium opacity peak may exist, the helium opacity is typically not as large as the iron opacity due to the low density. The turbulent pressure generated by the iron opacity zone completely changes the picture. Density drops much slower with radius when turbulent pressure support is significant as shown in Figure \ref{fig:compare3D_1D}. This will significantly increase the value of helium opacity due to the increased density. As demonstrated by \cite{Jiang2018},  the strong helium opacity can cause episodic mass loss from  massive stars with time-averaged mass loss rate $\sim 10^{-7}-10^{-5}\dot{M}_{\odot}/{\text{yr}}$. This can be an important mechanism for the enhanced mass loss rate  when massive stars evolve from the main sequence to the cooler side of the HR diagram.  The large mass loss rate may also cause the star to lose the outer layer of the envelope in a few years and thus have a larger effective temperature. This can potentially explain the effective temperature variations of luminous blue variables \cite{Jiang2018}. A similar idea was also developed based on a temperature dependent mass-loss prescription in 1D time-dependent hydrodynamic stellar evolution models to explain the S Doradus variability \cite{Grassitelli+2021}. If the mass loss rate is assumed to vary from $5\times 10^{-4}\dot{M}_{\odot}/{\text{yr}}$ to $1.26\times 10^{-3} \dot{M}_{\odot}/{\text{yr}}$ when the temperature at the sonic point changes from $2.5\times 10^4$ K to $2\times 10^4$ K, it can cause the envelope structure to change on the thermal timescale and cyclic variations of the stellar radii and effective temperatures in a way that is consistent with observations.


\subsection{Porosity}
\label{sec:porosity}

\begin{figure}[htp]
\centering
\includegraphics[width=1.0\hsize]{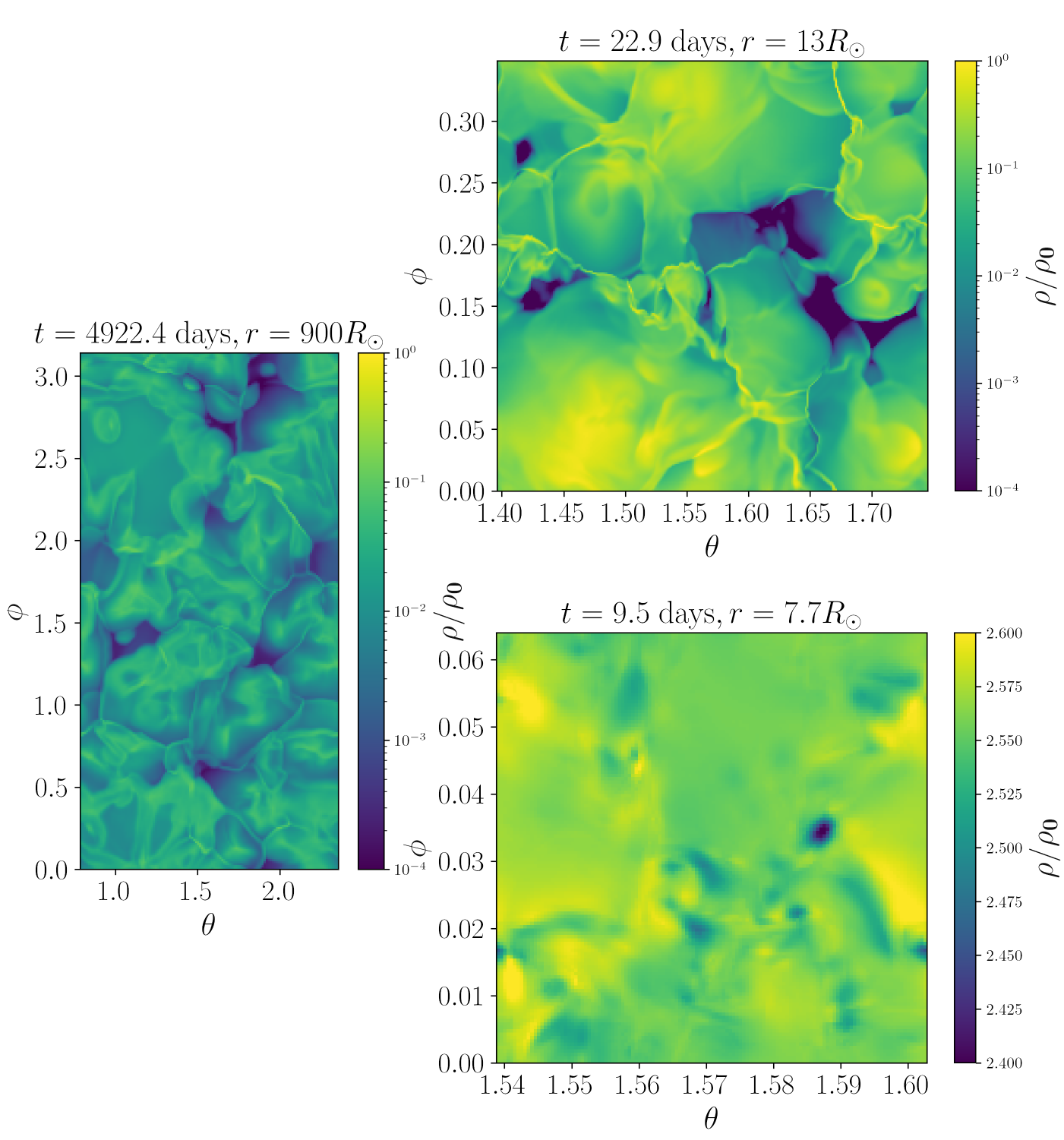}
\caption{Slices of density distribution at fixed radii near the photosphere from the three models $RSG2L4.9$ (left panel), $T35L5.2$ (top right panel) and $T42L5.0$ (right bottom panel). The time corresponding to each snapshot and radii we take the slices are shown in each panel.   }
	\label{fig:density2D}
\end{figure}

Convective turbulence in 3D will unavoidably cause temporal and spatial fluctuations of all the radiation hydrodynamical quantities to the photosphere, even though the surface region is supposed to be hydrostatic in 1D models. The normalized standard deviation of density at each radius is typically proportional to $\mathcal{M}^2$ \cite{Goldberg2022}, where $\mathcal{M}$ is the ratio between turbulent velocity and sound speed. Near the photosphere where photon diffusion is rapid, the fluctuation amplitude can be even larger than what is predicted by this scaling. Therefore, near the base of the convective zone where $\mathcal{M}$ is typically smaller than $0.1$ (Figure \ref{fig:turbulentpressure}), the fluctuation amplitude is only a few percent of the mean value. However, the relative fluctuations can become the order of unity near the photosphere. The characteristic size of turbulent eddies is then determined by the pressure scale height.  Slices of density at fixed radii near the photosphere for the three models $RSG2L4.9$, $T35L5.2$, and $T42L5.0$ are shown in Figure \ref{fig:density2D}. The $\theta$ and $\phi$ ranges correspond to the mesh size used for each simulation in the spherical polar coordinate, which covers at least several times $H_p/r$. For the zero-age main sequence $35M_{\odot}$ model $T42L5.0$, density fluctuation is smaller than $8\%$ of the mean value and the typical size of each filament is only $1\%$ of the stellar radius. This is in contrast to the other two models, where density can vary by several orders of magnitude at the same radius. In the high density region, optical depth per pressure scale height is still much larger than one, while in the low density region, it is already very optically thin. These two models also have turbulent velocities exceeding the sound speed near the photosphere. 

Other quantities, such as radiation flux, radiation energy density and opacity, also have similar fluctuations. More importantly, fluctuations of different quantities can be correlated, which can make the spatially averaged non-linear terms in 3D very different from the corresponding terms in 1D. One particularly important example is the radiation acceleration term. In the optically thick regime, the radiation pressure gradient is related to the co-moving radiation flux via the diffusion equation
\begin{eqnarray}
  \bfnabla P_r=-\frac{\rho\kappa_t}{c} \bF_{r,0}.
\end{eqnarray} 
After averaging the left and right hand sides for a fixed radius in 3D models, we get
\begin{eqnarray}
  \langle\bfnabla P_r\rangle=-\langle\frac{\rho\kappa_t}{c} \bF_{r,0}\rangle.  
\end{eqnarray}
In 1D models, if we treat $P_r,\rho,\kappa_t, F_{r,0}$ as corresponding shell averaged 3D values, the diffusion equation is typically taken to be
\begin{eqnarray}
\bfnabla \langle P_r\rangle=-\frac{1}{c}\langle \rho\rangle \langle \kappa_t\rangle \langle F_{r,0}\rangle.
\end{eqnarray}
However, due to the correlations between fluctuations of $\rho$ and $F_{r,0}$ in 3D, we can have $\langle \rho \kappa_t F_{r,0}\rangle\neq \langle \rho\rangle \langle \kappa_t\rangle \langle F_{r,0}\rangle$.
In particular, fluctuations of radiation flux $F_{r,0}$ can be anti-correlated with density fluctuations in the optically thick region so that $\langle \rho \kappa_t F_{r,0}\rangle < \langle \rho\rangle \langle \kappa_t\rangle \langle F_{r,0}\rangle$ inside the photosphere. This is called the porosity effect \cite{Shaviv1998} in the literature. Physically, this means that the low density region as shown in Figure \ref{fig:density2D} tends to have a larger co-moving radiation flux, while the high density filaments typically have smaller values of $F_{r,0}$. This is because rapid photon diffusion can keep the temperature at each radius roughly the same. Therefore, a larger optical depth between neighboring radial zones will correspond to a smaller diffusive flux for the same temperature gradient according to the diffusion equation. 
The ratio $\langle \rho \kappa_t F_{r,0}\rangle/\left(\langle\rho\rangle\langle\kappa_t\rangle\langle F_{r,0}\rangle \right) $ clearly depends on the 3D  nature of turbulence. In general, the larger amplitude density fluctuations can get, the more important this effect is. 

\begin{figure}
	\centering
	\includegraphics[width=0.8\linewidth]{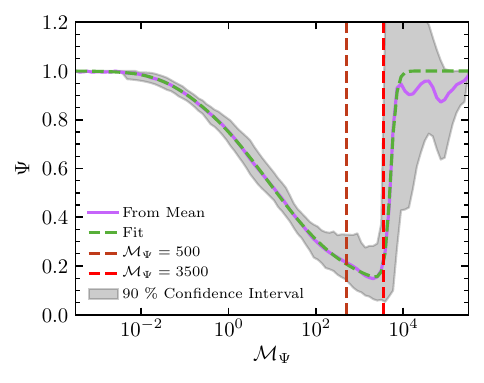}
	\caption{Dependence of $\Psi$ (equation \ref{eq:psi}) on $\mathcal{M}_{\Psi}$ (equation \ref{eq:mpsi}) from 3D simulations of various massive star envelopes, which is taken from Figure 3 of \cite{Schultz2020}. For each model, the horizontal axis is effectively the stellar radius increasing from left to the right. 
		The light purple line represents $\Psi$ as calculated from the mean of the variance distributions while the grey regions are the 90$\%$ confidence intervals from this mean. 
		The dashed green line is a fitting function to $\Psi$. 
		The two vertical dashed lines indicate the location where $\mathcal{M}_{\Psi}=500$ 
		and $\mathcal{M}_{\Psi}=3000$, where supersonic turbulence typically develops. Above this region, cross correlation between density and radiation flux fluctuations can invert. }
	\label{fig:psi}
\end{figure}

Properties of turbulence in the envelope of massive stars from the same models as shown in Figure \ref{fig:compare3D_1D} were studied by \cite{Schultz2020}. Fluctuations of all quantities are found to follow the log-normal probability distribution at a fixed radius. The impact of turbulence on $\bfnabla P_r$ in the optically thick regime is dominated by the anti-correlations between fluctuations of $\rho$ and $F_{r,0}$, which can be defined as 
\begin{eqnarray}
	\Psi\equiv \frac{\langle F_{r,0}\rho\rangle}{\langle F_{r,0}\rangle \langle \rho\rangle}.
	\label{eq:psi}
\end{eqnarray}
Notice that what matters is the fluctuation of co-moving frame radiative flux, not the lab-frame radiative flux, which can be quite different, particularly in the slow-diffusion optically thick regime. The difference between $\langle \rho \kappa_t F_{r,0}\rangle$ and $\langle \rho F_{r,0}\rangle\langle \kappa_t\rangle $ is found to be less than $1\%$ in the envelope and therefore the variation of opacity itself is neglected. The effect of turbulence on radiation force as quantified by $\Psi $ is typically larger with larger density fluctuations and local Eddington ratio. It can vary from 1 at the bottom of the convective zone to $\approx 0.1$ near the photosphere as shown in Figure \ref{fig:psi}. Changes of $\Psi$ as a function of radius and luminosity across different stellar models are found to be unified if the following pseudo Mach number is chosen to be the independent variable
\begin{eqnarray}
	\mathcal{M}_{\Psi}\equiv \frac{L}{4\pi a_r r^2 T^{4.5}}\left(\frac{\mu m_p}{k_B}\right)^{1/2}.
\label{eq:mpsi}
\end{eqnarray}
This is effectively the ratio between the energy transport speed for a given luminosity in the radiation pressure dominated regime and the local isothermal sound speed. \cite{Schultz2020} shows that the dependence of $\Psi$ on $\mathcal{M}_{\Psi}$ can be described by the same fitting formula (equation 12 of \cite{Schultz2020}, the dashed green line in Figure \ref{fig:psi}) for massive star models covering quite different mass, luminosity and metallicity. This is a promising way to incorporate this effect in 1D stellar evolution models. 

However, the turbulent properties change once they close to the photosphere. The cross-correlations between density and flux fluctuations also change sign as shown in Figure \ref{fig:psi}. More importantly, the diffusion equation and the above analysis are not valid anymore near and above the photosphere. It also does not guarantee that turbulence generated by different mechanisms, for example, the line-deshadowing instability (LDI, \cite{Owocki2015}), will have the same probability distribution in the optically thin part of the flow. 

Winds from massive stars are known to be clumpy \cite{Vink2022} and a formula has been developed to incorporate the effects of clumpy media or porosity in 1D models \cite{Grafner+2012}. One type of effort is trying to model the effect of density fluctuations on the opacity $\kappa_t$ itself while the transport effect (the corresponding variation of radiation flux) is neglected. 
If the flow is completely optically thin, it is assumed that all the mass is in a total number of $N$ clumps with density larger than the mean value by a factor of $N$ but they only occupy $1/N$ of the volume. If the opacity varies with the density as $\rho^2$, the mean opacity will be reduced by a factor of $N$, and thus the mean radiation acceleration is reduced by the same factor. When the clumps become optically thick, the mean opacity is thought to be reduced by another factor $\left[1-\exp(-\tau_c)\right]/\tau_c$ \cite{Owocki+2004}, where $\tau_c$ is the optical depth across each clump. This is designed so that when $\tau_c\rightarrow 0$, this additional factor is 1. When $\tau_c \gg 1$, the mean opacity is reduced by a factor of $1/\tau_c$. This formula is extended by \cite{Owocki+2018} to include a continuous distribution of density with the non-linear structures formed by LDI as an example. These formulae are unlikely to apply to the optically thick turbulence as found by the 3D simulations of massive star envelopes. The variation of opacity $\kappa_t$ itself with density plays a very minor role in the dynamics of the simulations, which is not surprising as the iron opacity peak is more sensitive to temperature but less sensitive to density. The change of radiation force as quantified by the parameter $\Psi$ can be thought of as a change of effective opacity with the same $\langle \rho \rangle$ and $\langle F_{r,0}\rangle $. However, as shown in Figure \ref{fig:psi}, $\Psi$ is always 1 when density fluctuation is very small but it can vary significantly with values larger or smaller than 1 in the optically thin region, which cannot be captured by the simple formula $\left[1-\exp(-\tau_c)\right]/\tau_c$.

Even if $\kappa_t$ is independent of density (such as electron scattering) or depends on density linearly, it has been known that density fluctuations can affect the transport of the photons and effectively reduce the opacity \cite{Shaviv1998,Oskinova+2004}. This is indeed the same effect as shown in Figure \ref{fig:psi}. It will be very interesting to see how this effect will vary for turbulence generated by different mechanisms.

\begin{figure}
	\centering
	\includegraphics[width=0.67\linewidth]{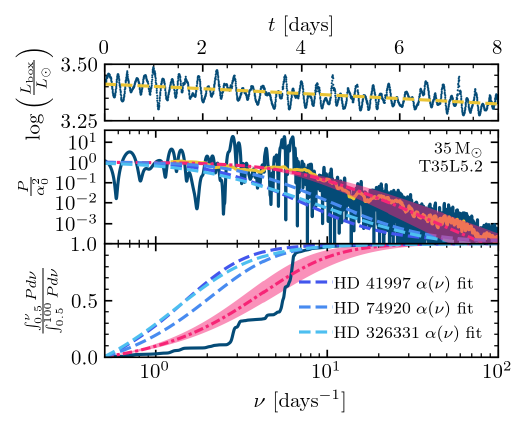}
	\caption{Lightcurve and power spectrum from the middle age main sequence model $T35L5.2$, which are taken from Figure 5 of \cite{Schultzetal2022}. 		
		Top: the dark blue points are the lightcurve for 8 days generated by the simulations while the gold dashed line is a first order polynomial fit to zero-mean the lightcurve before taking the power spectrum. 
		Middle: Power spectrum of the lightcurve from the top panel (solid dark blue line) compared to the normalized fit as used in three observed stars near the same location of HR diagram \cite{Bowman2020a} (dashed blue lines of different shades).
		The solid gold line shows the result of removing the periodic signals and linearly smoothing the power spectrum.
		The pink dot-dashed line denotes the fit to equation \ref{eq:slfv_fit} while the pink shaded region shows the $95\%$ confidence interval.
		Bottom: Cumulative power spectrum of all normalized power spectra on the middle panel normalize to be zero at the left limit ($\nu=0.5\,$days$^{-1}$) and one at the right edge.}
	\label{fig:slfv}
\end{figure}

\section{Observational Consequences}
\subsection{Impact on Photometric and Spectroscopic Variability}
One of the somewhat surprising observational properties of massive stars is the ubiquitous low amplitude temporal brightness variability as probed by space-based telescopes (e.g. TESS \cite{Ricker+2015}). All massive stars exhibit broad-band photometric variability up to $5\,$mmag ($\approx 0.5\%$) on timescales of hours to days \cite{Bowman2019b,Bowman2020a,Bowman2020b},  regardless of their spectral class, metallicity, or rotation rate. 
This is referred to as stochastic low frequency variability (SLFV) and the variability amplitude is typically larger for stars with larger luminosity and lower effective temperature.  

Different theories have been proposed to explain the origin of SLFV. One possible cause is internal gravity waves (IGWs) generated in the convective hydrogen burning cores of these massive stars, which then propagate through the radiative envelope and manifest on or near the stellar surface  \cite{Aerts2015,Bowman2019a,Edelmann+2019}. If this is true, SLFV can be a powerful tool to probe the internal structures of massive stars. However, the spectrum shape and variability amplitude of IGWs reaching the stellar photosphere depend on the propagating and damping properties of these waves. Recent studies suggest that with realistic stellar parameters \cite{Lecoanet2019,Anders+2023}, IGWs may not be able to explain the commonly observed SLFV for massive stars. Inhomogeneities from stellar winds combined with rotational effects as well as line-driven wind instabilities have also been proposed as possible explanations for SLFV \cite{David-Uraz2017,Simon-Diaz2018,Krticka2018,Krticka2021}.

Surface convection in massive star envelopes is another promising mechanism for producing SLFV. As shown in previous sections, turbulence driven by the iron opacity peak in 3D simulations is not just confined to the radial range where the opacity peak is located. The "quiet" radiative zone near the surface region typically found in 1D models does not exist. Instead, the photosphere region can have turbulent velocity $10-100$ ${\rm km/s}$, which can naturally cause fluctuations in the lightcurves. Figure \ref{fig:slfv} shows an example lightcurve generated by the simulation of a middle age main sequence $35M_{\odot}$ model described in \cite{Schultzetal2022}.
The lightcurve indeed shows stochastic variations with maximum contrast of luminosity reaching $\approx 5\%-10\%$ of the mean value. The power spectrum of the lightcurve shows a flat shape with frequency $\nu$ below a characteristic value $\nu_{\rm char}=7.2 $ day$^{-1}$, which is pretty consistent with $1/t_{\rm th}$, where 
$t_{\rm th}$ is the thermal time scale to the iron opacity location. For frequency above $\nu_{\rm char}$, the power spectrum decreases with increasing $\nu$. Observationally, SLFV is typically fitted with the following function shape
\begin{eqnarray}
	\alpha(\nu)=\frac{\alpha_0}{1+\left(\nu/\nu_{\rm char}\right)^{\gamma}}+C_W,
\label{eq:slfv_fit}
\end{eqnarray}
which reaches a constant value when $\nu\rightarrow 0$ and drops as a power law $\nu^{-\nu}$ for large frequency. Here $C_W$ is a white noise floor. The best-fitted values are $\alpha_0=0.0023\pm 0.0005$ and $\gamma=1.9\pm 0.2$ for this model. Since the simulation domain is a representative wedge of the whole star covering only $1/50$ of the full surface, the  variability amplitude of the integrated lightcurve from the whole star should be smaller by a factor of $\sqrt{50}$ if different wedges vary independently. This agrees with the observed values 
$\alpha_0=0.03\%-0.1\%$ and $\gamma=1.7-2.3$ very well for stars around the similar region of HR diagram \cite{Bowman2020a}, although the characteristic frequency $\nu_{\rm char}$ from this model is larger than the observed values for stars with similar luminosity and effective temperature by a factor of $\approx 2$.

The simulations predict a very clear dependence of SLFV on stellar parameters. The variability amplitude should get smaller when stars get closer to the zero-age main sequence, as the iron opacity peak moves closer to the surface with a smaller $\tau_{\rm Fe}$ and therefore the surface convection does not have enough time to affect the photon variability before they escape from the photosphere. This is confirmed by  a ZAMS $35M_{\odot}$ model shown in \cite{Schultzetal2022}, where the best-fitted $\alpha_0$ is only $3\times 10^{-6}$. The characteristic frequency $\nu_{\rm char}$ also gets larger as the thermal timescale to the convection zone is smaller. Stars with smaller mass should also have a lower variability amplitude 
as the Eddington ratio is reduced, although SLFV should still be observable for $13M_{\odot}$ as shown in \cite{Schultz+2023}. It is encouraging that all these trends are very consistent with the observed properties of SLFV reported by \cite{Bowman2020a}, although direct one-to-one comparison is still hard as the simulated model may not have the same stellar parameters as the observed ones.



\subsection{Impact on Spectroscopic Fitting of Massive Stars}
Rotation rates of massive stars are typically measured by fitting the observed shapes of spectral lines with theoretical expected broadening due to rotation \cite{Gray2005}. However, it is known that an extra line broadening mechanism in addition to rotation broadening is needed in many objects, in particular, the high luminosity giants and supergiants \cite{Slettebak1956,Conti1977,Howarth1997,Ryans+2002}.
Typically there are four main velocity components used to fit photospheric spectral lines \cite{Gray2005}.
Thermal broadening comes from the intrinsic Maxwell-Boltzmann velocity distribution of the ions, $v_{\rm therm}$, and is Gaussian in profile.
This is often combined with the intrinsic broadening, arising from the atomic physics governing the line-level transitions, to generate a Voigt profile for the spectral line. 
Projected rotational broadening, $v\sin i$, imparts a deep, steep-walled trench shape on stellar spectral lines as half the star is red-shifted while the other half is blue-shifted.
The other two velocity components are thought to come from the turbulent motions in the line forming regions, which are often needed to improve the fitting in different parts of the line profile. 
The microturbulent velocity denoted as $\xi$, is defined as an additional velocity impacting scales smaller than the emitting region and is added in quadrature to $v_{\rm therm}$ in the Gaussian broadening of spectral lines. Its typical value is smaller than the sound speed near the photosphere. As $\xi$ affects the equivalent width, it is typically quantified using Curve of Growth analyses of heavy elements for which $\xi \gg v_{\rm therm}$. The last velocity component is typically needed to broaden the wings of the spectral lines \cite{Gray2005}, which is called macroturbulence $v_{\rm macro}$ with values $\approx 20-80$ $\rm{km/s}$. This is thought to come from dynamics on scales larger than the emitting region with velocities that can exceed local sound speed. Though these velocity choices are very effective in fitting hot, massive star spectral lines \cite{Simon-Diaz2010, SimonDiaz2014,Simon-Diaz2017,Holgado+2022}, the origin of 
these large-scale turbulent velocities at the photosphere is unclear. Since the inferred macroturbulent velocity can be comparable to the measured rotational speed, systematic uncertainties can be introduced to the measured values of rotation depending on how the macrotubulent velocities are modeled. One example of an unexplained discrepancy is a strong positive correlation between $v\sin i$ and $v_{\rm macro}$ for massive stars ($M>20\,M_{\odot}$) across the main sequence \cite{Simon-Diaz2017}.


Surface convection with supersonic turbulent velocity at the photosphere is a natural way to broaden these lines. As a proof of principle, \cite{Schultz+2023b} post-processed the radiation hydrodynamic simulations discussed in previous sections with the Monte Carlo radiation transport code SEDONA \cite{Kasen+2006}, to quantify the broadening of individual photospheric lines due to the turbulent flow. Photons at the frequency of the OIII line are seeded below the photosphere for two 35 solar mass models at zero-age and terminal-age main sequences. After they propagate out of the photosphere by passing through the turbulent flow, the full width at half maximum (FWHM) of the line is $\approx 20$ km/s for the zero-age main sequence model, which is consistent with thermal broadening of the line. However, the line profile has a much broader tail compared with the commonly used Voigt line profile. FWHM is increased to $\approx 80$ km/s for the terminal age main sequence model as the turbulent velocity is increased near the photosphere. The line profile can also have significant time dependence on the timescale of a few days when the photosphere radius is variable due to episodic mass loss such as the 80 solar mass model shown in \cite{Jiang2018}. These preliminary calculations show that the 3D turbulent photosphere is very promising to cause the additional broadening of lines besides rotation. The mechanism also has a clear prediction on the amount of turbulent broadening as a function of stellar mass and evolution stages, which will vary in the same way as the amplitude of stochastic low frequency variability as they are caused by the same physical process. Adding rotation to the envelope models and quantifying the line shapes due to turbulent and rotation broadening together will be a natural way to reduce the systematic errors in the measurement of massive star rotation period.

\section{Future Directions}
Radiation force due to numerous lines in the optically thin region is known to play an important role in driving the mass loss of massive stars \cite{Owocki2015,Vink2022} and one important direction for future models of massive star envelopes is to include the contributions of the line forces in an appropriate and self-consistent way. 
 It is still too expensive to resolve the transport of individual lines or groups of lines so that the line force can be accounted for correctly in 3D radiation hydrodynamic simulations. Studies of line-driven winds typically adopt the CAK formula \cite{Castor1975}  based on the Sobolev approximation\cite{Sobolev1960}. It assumes that the stellar photons only interact with the wind over a narrow resonance layer with a width set by the Sobolev length $l_{\rm Sob}=v_{\rm thermal}/(dv/dr)$, where the velocity of the wind is assumed to be monotonic along radial direction $r$. The intrinsic width of the line is set by $v_{\rm thermal}$ as the stellar photosphere is thought to be static, which is not true given the turbulent envelope structures revealed by the 3D simulations. Then radiation acceleration is determined by the luminosity and a force multiplier, which is typically taken to be a power law of the optical depth across the Sobolev length and accounts for the contributions of radiation force due to all the lines. 

The same framework has been used in multi-dimensional radiation hydrodynamic simulations of Wolf-Rayet winds \cite{Moens+2022b}. Radiation acceleration at each spatial location is taken to be $\rho \kappa_{\rm sum} \bF_{r,0}/c$, where co-moving radiation flux $\bF_{r,0}$ is determined based the diffusion approximation using radiation energy density gradient and the total opacity $\kappa_{\rm sum}$. The opacity is taken to be the sum of Rosseland mean values and contribution from lines as
$\kappa_{\rm sum}=\kappa_R+\kappa_{\rm line}$, where $\kappa_{\rm line}$ is determined by the CAK formula based on the radial velocity gradient $|dv/dr|$ as the velocity may not be monotonic along the radial direction in 3D. Notice that $\kappa_{\rm line}$ is added to the opacity everywhere even though the flow is optically thick. 
This is an interesting attempt trying to couple the line forces with the 3D turbulence envelope together, although there are several concerns regarding this framework. As discussed in section \ref{sec:method}, Rosseland mean opacity is the appropriate opacity to determine the momentum coupling between radiation and gas if the co-moving frame flux is given by the diffusion equation for all the relevant frequency ranges. That also means when there is a  need to correct the Rosseland mean opacity due to lines that are not optically thick, diffusion approximation is also not a good approximation. Therefore, it is conceptually not self-consistent to use diffusion approximation to determine $\bF_{r,0}$ while correcting Rosseland mean opacity by adding $\kappa_{\rm line}$. It is also known that diffusion approximation can underestimate the turbulent velocity due to acceleration by the continuum radiation field for certain applications when the optical depth is in order of unity \cite{Davis2014}. Another big concern is the applicability of Sobolev approximation in 3D turbulent flow, particularly the use of $|dv/dr|$ to determine $\kappa_{\rm line}$. Since the transverse velocities can be comparable to the radial velocities (Figure \ref{fig:turbulentpressure})  and mass loss can be episodic, it is unclear whether local values of $|dv/dr|$ are enough to capture the effects of line force on the turbulent flow or not. Testing these assumptions and developing a more self-consistent model for line transport in 3D turbulent flow will be a crucial next step for massive star envelope models. 

The impact of rotation on the envelope structures as well as the impact of turbulence on the measurement of rotation need to be investigated in great detail. Rotation will provide additional support against gravity, which can potentially enhance the radiation-driven mass loss rate. Rotation could also modify the velocity distribution of the turbulent flow, which can affect the turbulent broadening of lines. Therefore, spectroscopic fitting based on models with rotation and surface convection together will be a great way to improve the current fitting procedure for stellar rotation measurement. 


It is known that most massive stars do not evolve alone and binary interactions can drastically modify the appearance, lifetime, and final fate of both stars \cite{Sana+2012,Sana+2013,Offner+2022}. However, when the mass transfer will occur in a binary system, stability and outcome of binary mass transfer depend crucially on the envelope 
structures of massive stars as well as the response of the envelope in a binary system. The turbulent pressure supported outer envelope will also have different thermodynamic properties compared with the traditional thermal pressure supported structures. Studying the dynamics of massive star envelopes in a binary potential using 3D radiation hydrodynamic simulations will be able to test various assumptions in binary stellar evolution models and improve our understanding of the structures and evolution of massive stars.


\vspace{6pt} 




\acknowledgments{The author thanks Matteo Cantiello for input on 1D stellar evolution models; Will Schultz, Jared Goldberg and Lars Bildsten 
	for analyzing the simulation results that are used in this review. 	The Center for Computational Astrophysics at the Flatiron Institute is supported by the Simons Foundation. }




	\newpage
\begin{adjustwidth}{-\extralength}{0cm}

\reftitle{References}



\bibliography{msenvelope}{}
\bibliographystyle{Definitions/mdpi}

%


\end{adjustwidth}

\end{document}